\documentclass[a4paper,11pt]{article}
\usepackage{jheppub}
\usepackage{eurosym}
\usepackage{amssymb}
\usepackage{amsfonts,cite}
\usepackage{amsmath}
\usepackage{graphicx}
\usepackage{multirow}
\usepackage{xcolor}
\usepackage{epsfig}
\usepackage{fancyhdr}

\newbox\mybox

\newcommand\fverb{\setbox\mybox=\hbox\bgroup\verb}
\newcommand\fverbdo{\egroup\medskip\noindent\fbox{\unhbox\mybox}\ }
\newcommand\fverbit{\egroup\item[\fbox{\unhbox\mybox}]}

\abstract{We investigate the Pais-Uhlenbeck (PU) model, a paradigmatic example of a higher time-derivative theory, by identifying the Lie symmetries of its associated fourth-order dynamical equation. Exploiting these symmetries in conjunction with the model’s Bi-Hamiltonian structure, we construct distinct Poisson bracket formulations that preserve the system's dynamics. Amongst other possibilities, this allow us to recast the PU model in a positive definite manner, adding another solution to the long-standing problem of ghost instabilities. Furthermore, we systematically explore a family of transformations that reduce the PU model to equivalent first-order, higher-dimensional systems. Finally we examine the impact on those transformations by adding interaction terms of potential form to the PU model and demonstrate how they usually break the Bi-Hamiltonian structure. Our approach yields a unified framework for interpreting and stabilizing higher time-derivative dynamics through a symmetry analysis in some parameter regime.}    

\title{Lie symmetries and ghost-free representations of the Pais-Uhlenbeck model}

 \author[a]{Alexander Felski,}
 \author[b]{Andreas Fring}
 \author[b]{and  Bethan Turner}

 \emailAdd{alexander.felski@mpl.mpg.de}
 \emailAdd{a.fring@city.ac.uk}	
 \emailAdd{Bethan.Turner.2@citystgeorges.ac.uk}

 \affiliation[a]{Max Planck Institute for the Science of Light, Staudtstraße 2,91058 Erlangen, 	Germany}
 \affiliation[b]{Department of Mathematics, City St George's, University of London,  Northampton Square, \\ London EC1V 0HB, UK}

\begin{document}
 	\maketitle
 	
 	\pagestyle{fancy}
 	\fancyhead{} 
 	\fancyhead[LE,RO]{\small\itshape  Lie symmetries and ghost-free representations of the Pais-Uhlenbeck model} 
 	
 	\renewcommand{\headrulewidth}{0.4pt}
 	
 	\section{Introduction}	

Higher time-derivative theories (HTDTs) have attracted significant attention across multiple domains of theoretical physics, particularly in the contexts of quantum field theory, modified theories of gravity, and non-local interactions \cite{Hawking,biswas2010towards,Salvio2,weldon98finite,mignemi1992black,plyush89mass,Mpl,dine1997comments,smilga17ultrav}. These theories typically involve Lagrangians with derivatives of order higher than one in time, which often emerge naturally in effective descriptions of fundamental interactions or in attempts to regularize ultraviolet divergences \cite{pais1950field,stelle77ren,grav1,grav2,grav3,modesto16super}. Despite their appealing theoretical features, HTDTs are notoriously plagued by issues of instability, most prominently, the emergence of ghost states, which signal the presence of unbounded Hamiltonians from below or non-normalisable states, thus threatening the physical viability of the theories \cite{weldon03quant,fring2024higher,fring2025quantisations}.

Among the simplest and most well-studied models within this class of HTDT is the Pais-Uhlenbeck (PU) oscillator \cite{pais1950field}. The PU model serves as a canonical example of a fourth-order differential system, encapsulating the essential challenges and features of HTDTs. It provides a fertile testing ground for investigating stability, quantization, and the role of symmetries in higher-derivative frameworks.

In recent years, there has been a growing interest in reformulating HTDTs in ways that circumvent the ghost problem \cite{ghostconst,salvio16quant,fakeons,bender2008no,raidal2017quantisation}, particularly through the identification of alternative Hamiltonian structures that preserve the dynamics but yield bounded spectra \cite{bolonek2005ham,dam2006,stephen2008ostro,most2010h,andrzejewski2014ham,and2014conflett,andrzejewski2014conf,elbistan2023various}. This often involves uncovering hidden symmetries or leveraging the Bi-Hamiltonian nature of the systems to construct equivalent descriptions with favourable properties.

Here we undertake a systematic exploration of the PU model using the framework of Lie symmetries. 

The Lie symmetries for the non-degenerate PU-equation have been studied before in \cite{nucci1,nucci2,nucci3}, where the following procedure was proposed: Identifying the Lie symmetries of the stand alone equation of motion, subsequently construct the Noether charges which are then combined in such a way that they lead to an autonomous Hamiltonian which has the same amount of Noether symmetries. In the case of the PU-model this has led the authors to find, as they claim, {\em the} Hamiltonian of the system. However, the procedure does neither guarantee the positive definiteness nor the uniqueness of the constructed Hamiltonian. Moreover, there have been examples of other positive definite Hamiltonians in the literature, challenging the idea that one can identify one particular Hamiltonian associated to the problem. Here we will demonstrate that there are in fact plenty of viable positive-definite Hamiltonians with different types of Poisson bracket structure that all preserve the dynamical flow of the system. 

Specifically, we identify the Lie symmetries of the fourth-order differential equation governing the PU oscillator and use these symmetries, in conjunction with the Bi-Hamiltonian structure, to construct Poisson brackets that preserve the system’s dynamics. These constructions enable us to systematically recast the PU model in many alternative variants most importantly including some that are manifestly positive definite, thereby eliminating the ghost issue in certain regimes. Furthermore, we develop and classify a family of transformations that map the higher-derivative PU dynamics into lower-order but higher-dimensional formulations. These reformulations not only retain the physical content of the original model but also open up new avenues for both classical and quantum analyses.

Our manuscript is organised as follows:  In section 2, we identify the Lie symmetries of the PU model and construct its Bi-Hamiltonian structure, including associated Poisson brackets and conserved quantities. We present a study of the flow of the classical solution under the action of the full Lie group. Section 3 presents a systematic class of transformations that map the PU model to two-dimensional first-order systems, deriving corresponding Hamiltonians and identifying those with positive-definite formulations. Particular emphasis is placed on the conditions under which the transformed models preserve the original flow and Poisson structures. In section 4 we examine the impact of potential interaction terms, showing how they generally break the Bi-Hamiltonian structure. Our conclusions are stated in section 5 with a summary and discussion of future directions.

\section{Lie symmetries, flows and Poisson bracket structures for the PU model}

In order to set the stage, we start by recalling some well-known and less well-known facts about the PU model \cite{pais1950field}. It originates from the Lagrangian density that crucially involves a second order time-derivative $\ddot{q}$ of the coordinate $q$, which is why it is widely regarded as the simplest prototype HTDT
\begin{equation}
	{\cal L}_{\text{PU}}(q,\dot{q} , \ddot{q})=\frac{1}{2}  \ddot{q}^2  -\frac{\alpha}{2}  \dot{q}^2 + \frac{\beta}{2} q^2 , \qquad \alpha,\beta \in \mathbb{R} .  \label{LPU}
\end{equation}
The Euler-Lagrange equation resulting from $	{\cal L}_{\text{PU}}(q,\dot{q} , \ddot{q})$ is the fourth order PU oscillator equation  
\begin{equation}
	\frac{d^2}{dt^2}   \frac{	\delta {\cal L}}{\delta \ddot{q}} - \frac{d}{dt} \frac{\delta  {\cal L} }{\delta \dot{q}} + \frac{\delta  {\cal L}}{\delta q}  =0, \qquad \Rightarrow \qquad \ddddot{q} + \alpha \ddot{q} + \beta q =0. \label{equm1}
\end{equation}
Defining the vector $\vec{q} = \{ q , \dot{q},\ddot{q}, \dddot{q}     \} \in \mathbb{R}^4$, the dynamical equations of the PU oscillator can be cast into the form
\begin{equation}
	\frac{d\vec{q}}{dt} = \vec{V}(\vec{q}),   \qquad  V= \sum_{i=1}^4 v_i \partial_{q_i}   = \dot{q} \partial_q +  \ddot{q} \partial_{\dot{q}} 
	+  \dddot{q} \partial_{\ddot{q}} - \left( \alpha \ddot{q} + \beta q \right) \partial_{ \dddot{q} },  \label{1flow}
\end{equation}
where the first order linear differential operator $V$ acts like a vector field when interpreted as a derivation on $C^{\infty}(\mathbb{R}^4)$.
\subsection{Lie symmetries of the PU oscillator equation}
Next we identify the Lie symmetries of the dynamical system in (\ref{1flow}) in the usual way, see e.g. \cite{hydonsymm}. Infinitesimally transforming $q_i \rightarrow \tilde{q}_i = q_i + \varepsilon \xi_i$, $\varepsilon \ll 1$ the component version of (\ref{1flow}) transforms to
\begin{equation}
     \frac{d \tilde{q}_i}{dt} = V_i(\tilde{q}) - \delta V_i(\tilde{q}), \qquad \text{with} \,\,\,\,  \delta V_i(\tilde{q}) = \varepsilon \left[X,V\right]_i(\tilde{q}) = \varepsilon  \sum_{j=1}^4   \left(  \xi_j \frac{\partial v_i}{ \partial q_j} - v_j \frac{\partial \xi_i}{ \partial q_j} \right),
\end{equation}
where the Lie algebraic generators are
\begin{equation}
   X = \sum_{i=1}^4 \xi_i \partial_{q_i}  .
\end{equation}
Using a generic Ansatz for the vector field $\xi_i$ as being linear in the coordinates $q_i$, we find four linearly independent solutions for the Lie bracket to vanish  
\begin{eqnarray}
	X_1 &=&  \dot{q} \partial_q +  \ddot{q} \partial_{\dot{q}} 
	+  \dddot{q} \partial_{\ddot{q}} - \left( \alpha \ddot{q} + \beta q \right) \partial_{ \dddot{q} }, \label{Lie1} \\
		X_2 &=&   \frac{1}{2} \left[  q \partial_q +  \dot{q} \partial_{\dot{q}} +
		  \ddot{q} \partial_{\ddot{q}} + \dddot{q}   \partial_{ \dddot{q} }  \right],      \label{Lie2}        \\
		X_3 &=&  \frac{1}{2} \left[ \ddot{q} \partial_q +  \dddot{q}  \partial_{\dot{q}} 
	-  ( \alpha \ddot{q} + \beta q    ) \partial_{\ddot{q}}    -  ( \alpha \dddot{q} + \beta \dot{q}    )   \partial_{ \dddot{q} }   \right],      \label{Lie3}        \\
		X_4 &=&  ( \alpha \dot{q} +\dddot{q}) \partial_q - \beta q  \partial_{\dot{q}} 
		- \beta \dot{q} \partial_{\ddot{q}} - \beta  \ddot{q}   \partial_{ \dddot{q} }.      \label{Lie4}
\end{eqnarray}
Here trivially $X_1$ is simply the dynamical vector field $V$,  $X_2$ is the Euler symmetry operator and  $X_3$, $X_4$ are less obvious. Their precise nature will be discussed below. We also note that the algebra formed by these generators is Abelian as all of them mutually commute, i.e $[X_i,X_j]=0$, for $i,j=1,\ldots,4$. 
\subsection{Standard Poisson bracket structure}
Starting from the flow of the dynamical vector field let us now identify a Hamiltonian together with its associated Poisson bracket structure by solving
\begin{equation}
	\vec{v} = \left\{  \vec{q}, H    \right\}       = J \nabla H ,    \label{flow1}
\end{equation}
for the Poisson tensor $J$ and the Hamiltonian $H$ with the vector field $\vec{v}$ as defined in (\ref{1flow}). From a generic expression as a quadratic form over the variables $\left\{ q, \dot{q}, \ddot{q} , \dddot{q}  \right\}$, we find the standard PU-Hamiltonian 
\begin{equation}
	{\cal H}_1 \left( q, \dot{q}, \ddot{q} , \dddot{q}  \right)= \frac{1}{2} \ddot{q}^2 - \frac{1}{2} \alpha \dot{q}^2   - \frac{1}{2} \beta q^2  -\dot{q} \dddot{q} , \label{Ham1}
\end{equation}
with Poisson bracket tensor
\begin{equation}
	J_1 = \left(
	\begin{array}{cccc}
		0 & 0 & 0 & -1 \\
		0 & 0 & 1 & 0 \\
		0 & -1 & 0 & \alpha  \\
		1 & 0 & -\alpha  & 0 \\
	\end{array}
	\right).
\end{equation}
Explicitly, the only non-vanishing brackets amongst the set of variables $\{ q, \dot{q}, \ddot{q} , \dddot{q}  \}$ are
\begin{equation}
	\{ \dot{q}, \ddot{q}  \}_1=1, \qquad   	\{ \dddot{q}, q  \}_1=1,  \qquad   \{ \ddot{q}, \dddot{q}  \}_1=\alpha .
\end{equation}
Thus, in general for any two functions $F(q, \dot{q}, \ddot{q} , \dddot{q})$ and $G(q, \dot{q}, \ddot{q} , \dddot{q})$, we have the induced Poisson bracket
\begin{equation}
	\{ F,G \}_1:=  \nabla F J_1 \nabla G =    \frac{\partial F}{\partial \dddot{q}}  \frac{\partial G}{\partial q} -  \frac{\partial G}{\partial \dddot{q}}  \frac{\partial F}{\partial q}  + 
	\frac{\partial F}{\partial \dot{q}}  \frac{\partial G}{\partial \ddot{q}} - \frac{\partial G}{\partial \dot{q}}  \frac{\partial F}{\partial \ddot{q}}  + \alpha \left(    
	\frac{\partial F}{\partial \ddot{q}}  \frac{\partial G}{\partial \dddot{q}} - \frac{\partial G}{\partial \ddot{q}}  \frac{\partial F}{\partial \dddot{q}}   \right) .   \label{Poi1} 
\end{equation}
Alternatively, we can  obtain $	{\cal H}_1$ with its Poisson bracket structure following Ostrogradky's approach \cite{ostrogradsky1850memoire}. The second-order Lagrangian (\ref{LPU}) can be converted into a two-dimensional first-order Hamiltonian system by defining the canonical variables:
\begin{equation}
	q_1 = q,    \quad   q_2 = \dot{q},   \quad          \pi_1 = \frac{\partial {\cal L} }{\partial \dot{q}} - \frac{d}{dt}  \frac{\partial {\cal L}}{\partial \ddot{q}}  = - \dddot{q} - \alpha \dot{q},  \quad
	    \pi_2 = \frac{\partial {\cal L}}{\partial \ddot{q}}  = \ddot{q} .   \label{equm3}
\end{equation}	
The corresponding Hamiltonian then results to 
\begin{equation}
	{\cal H}_{\text{PU}}(q_1,q_2,\pi_1,\pi_2)= \pi_1 \dot{q}_1 + \pi_2  \tilde{q} - {\cal L} (q_1,q_2, \tilde{q} ).
\end{equation}
where $\tilde{q}$ is to be understood as  $\tilde{q}   \equiv \ddot{q} =f(q_1,q_2,\pi_2)$. Thus, with 
\begin{equation}
	{\cal L}_{\text{PU}}(q_1,q_2,\tilde{q})= \frac{1}{2} \pi_2^2 -  \frac{\alpha}{2} q_2^2 + \frac{\beta}{2} q_1^2,
\end{equation}
we obtain the two-dimensional system
\begin{equation}
	{\cal H}_{\text{PU}}(q_1,q_2,\pi_1,\pi_2)= \pi_1 q_2 + \frac{1}{2} \pi_2^2 + \frac{\alpha}{2} q_2^2 - \frac{\beta}{2} q_1^2 . 
\end{equation}
Converting back to the $\vec{q}$-variables this Hamiltonian equals ${\cal H}_1$ in (\ref{Ham1}). This representation makes it possible to analyse the model using standard Hamiltonian techniques and investigate alternative formulations that preserve or modify its structure. For the higher order transformation (\ref{equm3}) the canonical Poisson bracket structure
\begin{equation}
	\{ F,G \}_c:= \sum_{k=1}^2  \left(  \frac{\partial F}{\partial q_k}  \frac{\partial G}{\partial \pi_k} -   \frac{\partial G}{\partial q_k}  \frac{\partial F}{\partial \pi_k}         \right),
\end{equation}
with $\{ q_i , \pi_j    \}_c = \delta_{ij}$, consistently reproduces Hamilton's equations for the dynamics of the system
\begin{eqnarray}
	\dot{q}_1 &=& \frac{\partial {\cal H}_{\text{PU}}  }{\partial \pi_1} =q_2 = \{ q_1,  {\cal H}_{\text{PU}}    \}_c , \qquad   \dot{\pi}_1 = -  \frac{\partial {\cal H}_{\text{PU}}  }{\partial q_1}  = \beta q_1  = \{ \pi_1,  {\cal H}_{\text{PU}}    \}_c ,  \label{dyn1}  \\
	\dot{q}_2 &=& \frac{\partial {\cal H}_{\text{PU}}  }{\partial \pi_2} =\pi_2 = \{ q_2,  {\cal H}_{\text{PU}}    \}_c , \qquad   \dot{\pi}_2 = -  \frac{\partial {\cal H}_{\text{PU}}  }{\partial q_2}  = - \alpha q_2 -\pi_1 = \{ \pi_2,  {\cal H}_{\text{PU}}   \label{dyn2}    \}_c . \qquad
\end{eqnarray}	
Transforming the $	\{ F,G \}_c$ Poisson bracket recovers precisely the previously obtained brackets $	\{ F,G \}_1$.
\subsection{Bi-Hamiltonian structure}
On the other hand it is known that the PU model is a Bi-Hamiltonian system, see e.g. \cite{dam2006}, meaning that exactly the same dynamics can be produced from two different Hamiltonians equipped with different Poisson brackets. Indeed,  there is a second solution to the dynamical flow equation (\ref{flow1}) \cite{dam2006}
\begin{equation}
	{\cal H}_2\left( q, \dot{q}, \ddot{q} , \dddot{q}  \right)= \frac{1}{2}  \beta \dot{q}^2 - \frac{1}{2} \alpha \ddot{q}^2   - \frac{1}{2}  \dddot{q}^2 -\beta q \ddot{q} ,  \label{Ham2}
\end{equation}
with 
\begin{equation}
	J_2= \left(
	\begin{array}{cccc}
		0 & \frac{1}{\beta } & 0 & 0 \\
		-\frac{1}{\beta } & 0 & 0 & 0 \\
		0 & 0 & 0 & -1 \\
		0 & 0 & 1 & 0 \\
	\end{array}
	\right),
\end{equation}
i.e. now the only non-vanishing brackets amongst the set of variables $\{ q, \dot{q}, \ddot{q} , \dddot{q}  \}$ are
\begin{equation}
	\{ q, \dot{q}  \}_2=\frac{1}{\beta}, \qquad     \{ \ddot{q}, \dddot{q}  \}_2=-1 .
\end{equation}
For two arbitrary functions $F(q, \dot{q}, \ddot{q} , \dddot{q})$ and $G(q, \dot{q}, \ddot{q} , \dddot{q})$,  the second Poisson brackets are therefore
\begin{equation}
	\{ F,G \}_2:=  \nabla F J_2 \nabla G = \frac{1}{\beta}   \left(  \frac{\partial F}{\partial q}  \frac{\partial G}{\partial \dot{q}}  -  \frac{\partial G}{\partial q}  \frac{\partial F}{\partial \dot{q}} \right) -
	\frac{\partial F}{\partial \ddot{q}}  \frac{\partial G}{\partial \dddot{q}} 
	+ \frac{\partial G}{\partial \ddot{q}}  \frac{\partial F}{\partial \dddot{q}} .  \label{Poi2} 
\end{equation}
Thus, we have 
\begin{equation}
J_2 \nabla 	{\cal H}_2 = J_1 \nabla 	{\cal H}_1. \label{j1j2ini}
\end{equation}
As is well-known for Bi-Hamiltonian systems, see e.g. \cite{Das,Magri}, these relations can be iterated as
\begin{equation}
	J_2 \nabla 	{\cal H}_{n+1} = J_1 \nabla 	{\cal H}_n, \qquad n= 1,2, \ldots,   \label{recn}
\end{equation}
where by construction all ${\cal H}_{n+1} $ are conserved quantities. This is easily seen, for instance for ${\cal H}_{3} $ we have
\begin{equation}
     \dot{ {\cal H}}_{3}= 	\left\{ 	{\cal H}_{3} ,	{\cal H}_{1}      \right\}_1 = \nabla	{\cal H}_{3} J_1 \nabla	{\cal H}_{1} =  \nabla	{\cal H}_{3} J_2 \nabla	{\cal H}_{2} = -  \nabla	{\cal H}_{2} J_2 \nabla	{\cal H}_{3} =
     -  \nabla	{\cal H}_{2} J_1 \nabla	{\cal H}_{2}  =0.
\end{equation}
Besides being conserved, all higher Hamiltonians are also in involution, i.e. $\{ {\cal H}_{i} , {\cal H}_{j}    \}_1=\{ {\cal H}_{i} , {\cal H}_{j}    \}_2 =0$. Using (\ref{recn}) one can simply iterate this argument to derive $ \dot{ {\cal H}}_{4}=0$ etc. Solving (\ref{recn}) recursively we find the conserved quantities 
\begin{eqnarray}
	  {\cal H}_{3} &=& \int J_2^{-1} J_1  \nabla {\cal H}_{2}  d \vec{q} =- \alpha  {\cal H}_{2} - \beta  {\cal H}_{1} , \\ 
      {\cal H}_{4} &=&  \int J_2^{-1} J_1   \nabla  {\cal H}_{3}  d \vec{q} = \left(\alpha^2 -\beta \right) {\cal H}_{2} + \alpha \beta  {\cal H}_{1}   , \\
      &\vdots& \notag
\end{eqnarray}
The Lie symmetries (\ref{Lie1})-(\ref{Lie4}) act as follows on those Hamiltonians 
\begin{equation}
	X_1({\cal H}_{i}) =  0, \quad  X_2({\cal H}_{i}) =  {\cal H}_{i}, \quad  X_3({\cal H}_{i}) = {\cal H}_{i+1}, \quad
	X_4({\cal H}_{i}) = 0,   \quad i=1,2, \ldots. \label{x2hi}
\end{equation}
Evidently, the dynamical flow $X_1$ leaves all four Hamiltonians invariant. From (\ref{x2hi}) we see that the Lie derivative $X_2$ can be identified as the dilation operator that rescales all Hamiltonians. 

While the Lie derivative $X_3$ is a dynamical symmetry it does not preserve the Hamiltonian. Instead it maps to the next higher charge. This feature allows us to utilise $X_3$ to construct the entire hierarchy of Hamiltonians from its consecutive actions. We find 
\begin{equation}
 X_3^k({\cal H}_{1}) = {\cal H}_{k+1} = \beta P_{k-1}  {\cal H}_{1}  + \frac{1}{\alpha} \left(  P_{k+1} +   \beta P_{k-1}   \right)  {\cal H}_{2}, \qquad k \in \mathbb{N},
\end{equation}
where we defined the polynomials
\begin{equation}
   P_n := \sum_{k=1}^{\lfloor \frac{n-1}{2} +1\rfloor }  c_k^n  \alpha^{n+1-2k} \beta^{k-1}, \quad \text{with} \,\,
     c_k^n = \frac{(-1)^{n+k+1}}{(k-1)!}  \prod_{\ell=k}^{2k-2} (n-\ell) .
\end{equation}
In the upper limit of the sum we used  the floor function ${\lfloor  x \rfloor}  := \max(n \in \mathbb{Z} \vert n \leq x  )  $ that gives the greatest integer less than or equal to $x$. Explicitly the first polynomials are 
\begin{equation}
	P_0=0, \,\, 	P_1=-1, \,\,   	P_2=\alpha, \,\,   	P_3=\beta - \alpha^2, \,\,  	P_4= \alpha^3 - \alpha \beta, \,\, 
	P_5= -\alpha^4 + 3 \alpha^2 \beta - \beta^2, \,\, \ldots 
\end{equation}
Thus, the iterated equation (\ref{recn}) is then simply obtained by acting with $X_3^n$ on relation (\ref{j1j2ini}). 

Interestingly, $X_4$ is not only a symmetry of the flow but also of the Hamiltonian ${\cal H}_{1}$ and  ${\cal H}_{2}$, i.e. it is a geometric symmetry of the entire phase space structure. In fact we can find two new Hamiltonians that generate the dynamical flow of $X_4$ in form of a new Bi-Hamiltonian structure. Defining the two Hamiltonians
\begin{equation}
	   \bar{{\cal H}}_{1} = \alpha  {\cal H}_{1} +  {\cal H}_{2}   , \qquad \text{and}  \qquad
	      \bar{{\cal H}}_{2} = -\beta  {\cal H}_{1}  
\end{equation}
we have the Bi-Hamiltonian structure
\begin{equation}
	X_4 (\vec{q}) = J_2 \nabla   \bar{{\cal H}}_{2} = J_1 \nabla   \bar{{\cal H}}_{1}  .
\end{equation}
We notice that $J_1$ and $J_2$ are not necessarily associated to either ${\cal H}_{1}$ and ${\cal H}_{2}$, respectively. It is therefore natural to consider the flow of the combined system. Taking the linear combination of both Hamiltonians  $ {\cal H}_{1}$ and  $ {\cal H}_{2}$ and a linear combination of the Poisson tensors $J_1$ and $J_2$, we define
\begin{equation}
	\bar{J} :=  c_1 J_1 +  c_2 J_2, \qquad \text{and} \qquad    \bar{{\cal H}} :=   c_3 {\cal H}_{1} +  c_4 {\cal H}_{2}. \label{barJH}
\end{equation}
From these expressions we compute 
\begin{equation}
	 \bar{J} \nabla \bar{{\cal H}} = \left( c_1  c_3 +  c_2  c_4 - c_1  c_4 \alpha \right) V(\vec{q})  +
	 \left(  c_1  c_4 -  c_2  c_3 \beta^{-1}   \right)   X_4(\vec{q}) .    \label{Lieconst}
\end{equation}
Thus, setting the coefficient factor in the first term to 1 and in the second to 0, we have 
\begin{equation}
	 \frac{d \vec{q}}{ dt} =  \bar{J} \nabla \bar{{\cal H}} = V(\vec{q}), \quad \text{for} \,\,
	  c_3 = \frac{ c_1 \omega_1^2 \omega_1^2}{ ( c_2 -  c_1 \omega_1^2)( c_2 -  c_1 \omega_2^2)  } ,   \quad
	  c_4 = \frac{ c_2 }{ ( c_2 -  c_1 \omega_1^2)( c_2 -  c_1 \omega_2^2)  }  .    \label{a1234}
\end{equation}
We used here the standard parametrisation $\alpha = \omega_1^2 + \omega_2^2$, $ \beta = \omega_1^2  \omega_2^2$, motivated by the classical solutions, see below (\ref{solndeg}), (\ref{soldeg}). Then we may re-write $\bar{{\cal H}} $ as a combination of positive definite terms, see \cite{dam2006}, with certain prefactors 
\begin{equation}
\bar{{\cal H}} = H_{12}+H_{21}, \qquad  H_{ij}=\frac{\omega_i^2}{2( c_1 \omega_i^2 -  c_2)(\omega_i^2 -\omega_j^2  )}
\left[  \left(  \dddot{q} + \omega_j^2 \dot{q}    \right)^2  +  \omega_i^2  \left(  \ddot{q} + \omega_j^2 q    \right)^2      \right] .   \label{H12H21}
\end{equation}
This means $\bar{{\cal H}} $ is positive definite when
\begin{equation}
              ( c_1 \omega_1^2 -  c_2)(\omega_1^2 -\omega_2^2  )  >0, \quad   \land \quad
              ( c_1 \omega_2^2 -  c_2)(\omega_2^2 -\omega_1^2  )  >0.   \label{PDC}
\end{equation}
There are clearly solutions for $ c_1$ and $ c_2$ for any order of the frequencies in the nondegenerate case. However, there is no solution with either $ c_1=0$ or $ c_2=0$. In other words, one can have a positive definite Hamiltonian that preserves the flow of the PU-oscillator, but only with an altered Poisson bracket structure as stated in (\ref{barJH}). We will identify such a structure below.

\subsection{Flows of the classical solutions}

Finally in this subsection, we also report the full flows generated by the Lie symmetries on the classical solutions which are known to be qualitatively of very distinct type depending on the parameter regime of the PU-equation (\ref{equm1}). The general real solutions are purely oscillatory 
\begin{equation}
q(t) = A_1 \sin( \omega _1 t) + A_2 \cos( \omega _1 t) + B_1 \sin( \omega _2 t) + B_2 \cos( \omega _2 t) \label{solndeg}
\end{equation}
in the nondegenerate case $\omega _1  \neq \omega _2 $, and oscillatory as well asymptotically divergent
\begin{equation}
	q(t) = A_1 \sin( \omega  t) + A_2 \cos( \omega  t) + B_1 t \sin( \omega  t) + B_2 t \cos( \omega  t) \label{soldeg}
\end{equation}
in the degenerate case when $\omega _1 =\omega _2 =: \omega $. 

Next we compute the full group action on the solution generated by the Lie symmetries, i.e. we associate to each solution a new point in phase space, characterised by the flow parameter $s$ via $\vec{q} \mapsto \phi_s(\vec{q} ) $, which means we need to solve
\begin{equation}
	  \frac{d}{ds}   \phi_s(\vec{q} )  = X(\phi_s(\vec{q} )), \qquad \text{with} \,\,\,  \phi_0(\vec{q}) = \vec{q}(t) .
\end{equation}
\underline{$X_2$-flow:} Given the representation (\ref{Lie2}) for the symmetry generated by $X_2$, we simply have to integrate the four decoupled equations
\begin{equation}
    \frac{d 	\phi_i^{(2)}(s) }{ds} = \frac{1}{2}	\phi_i^{(2)}(s)   \qquad \text{with} \,\,\,  \phi_i^{(2)}(0) = q_i(t), \quad i=1,2,3,4  .
\end{equation}
These are easily solved to 
\begin{equation}
	\phi_i^{(2)}(t,s)  = e^{s/2} q_i(t) ,
\end{equation}
i.e. the group action resulting from $X_2$ simply rescales any solution. 

\noindent \underline{$X_3$-flow:}  For the symmetry generated by $X_3$ with representation (\ref{Lie3}) the flow is obtained from the coupled equations
\begin{eqnarray}
	\frac{d 	\phi_1^{(3)} }{ds} &=& \frac{1}{2}	\phi_3^{(3)}  , \qquad  \qquad  \qquad  \qquad \qquad  \qquad \, 
		\frac{d \phi_2^{(3)} }{ds}	 = \frac{1}{2}	\phi_4^{(3)} , \quad\\
			\frac{d \phi_3^{(3)} }{ds}	 &=& - \frac{1}{2} \left[   \omega_1^2 \omega_2^2   \phi_1^{(3)} + ( \omega_1^2 + \omega_2^2 ) \phi_3^{(3)}  \right],
			 \quad  
			\frac{d 	\phi_4^{(3)} }{ds} =   - \frac{1}{2} \left[  \omega_1^2  \omega_2^2 \phi_2^{(3)}   +( \omega_1^2 + \omega_2^2) \phi_4^{(3)}  \right] \!  . \,\, \qquad \,\,
\end{eqnarray}
Given the initial conditions (\ref{solndeg}) in the nondegenerate case, we solve these equations to
\begin{equation}
		\phi_1^{(3)}(t,s)  = e^{-\frac{s}{2}  \omega _1^2} \left[ A_1\sin( \omega _1 t) + A_2 \cos( \omega _1 t) \right]+ e^{-\frac{s}{2}  \omega _2^2} \left[B_1 \sin( \omega _2 t) + B_2 \cos( \omega _2 t) \right] ,
\end{equation}
with $	\phi_2^{(3)}(t,s) = \partial_t   \phi_1^{(3)}(t,s)  $,  $	\phi_3^{(3)}(t,s) = \partial_t^2   \phi_1^{(3)}(t,s)  $ and $	\phi_4^{(3)}(t,s) = \partial_t^3   \phi_1^{(3)}(t,s)  $. 

\noindent Instead, for the degenerate case with initial conditions (\ref{soldeg}) we obtain 
\begin{equation}
	\phi_1^{(3)}(t,s)  = e^{-\frac{s \omega ^2}{2}} \left\{   \left[ A_1-B_2 s \omega +B_1 t\right]  \sin (t \omega )   + \left[ A_2+B_1 s \omega +B_2 t\right] \cos (t \omega )   \right\}  ,
\end{equation}
again with $	\phi_2^{(3)}(t,s) = \partial_t   \phi_1^{(3)}(t,s)  $,  $	\phi_3^{(3)}(t,s) = \partial_t^2   \phi_1^{(3)}(t,s)  $ and $	\phi_4^{(3)}(t,s) = \partial_t^3   \phi_1^{(3)}(t,s)  $. 

Thus, the group action resulting from $X_3$ rescales the solutions depending on the frequencies in the nondegenerate case and induces an additional shift in the degenerate case.

\noindent \underline{$X_4$-flow:}  For the symmetry generated by $X_4$ with representation (\ref{Lie4}) the flow is obtained from the coupled equations
\begin{eqnarray}
	\frac{d \phi_1^{(4)} }{ds}  &=&   	\phi_4^{(4)}  + (\omega_1^2 + \omega_2^2) \phi_2^{(4)},  \qquad
	\frac{d \phi_2^{(4)} }{ds} =  - \omega_1^2 \omega_2^2  \phi_1^{(4)},    \,\, \\
	\frac{d \phi_3^{(4)} }{ds} &=&  - \omega_1^2 \omega_2^2  \phi_2^{(4)},   \qquad  \qquad  \qquad \,
	\frac{d \phi_4^{(4)} }{ds}	 =  - \omega_1^2 \omega_2^2  \phi_3^{(4)}.
\end{eqnarray}
Given the initial conditions (\ref{solndeg}) in the nondegenerate case, we solve these equations to
\begin{eqnarray}
		\phi_1^{(4)}(t,s)  &=& A_1 \sin \left[\omega _1 \left(t+ s \omega _2^2\right)\right]
	+A_2 \cos \left[\omega _1 \left(t+s \omega _2^2\right)\right]
	+B_1 \sin \left[\omega _2 \left(t+s \omega_1^2\right)\right]   \\
	&&
	+B_2 \cos \left[\omega _2 \left(t+s \omega _1^2\right)\right] ,\notag
\end{eqnarray}
with $	\phi_2^{(4)}(t,s) = \partial_t   \phi_1^{(4)}(t,s)  $,  $	\phi_3^{(4)}(t,s) = \partial_t^2   \phi_1^{(4)}(t,s)  $ and $	\phi_4^{(4)}(t,s) = \partial_t^3   \phi_1^{(4)}(t,s)  $. 

\noindent For the degenerate case with initial conditions (\ref{soldeg}) we obtain 
\begin{equation}
	\phi_1^{(4)}(t,s)  = \left[ A_1+B_1 \left(t-s \omega ^2\right)\right] \sin \left[\omega  \left(t+s \omega ^2\right)\right] 
	+  \left[ A_2+B_2 \left(t-s \omega ^2\right)\right] \cos \left[ \omega  \left(t+s \omega ^2\right)\right],
\end{equation}
again with $	\phi_2^{(4)}(t,s) = \partial_t   \phi_1^{(4)}(t,s)  $,  $	\phi_3^{(4)}(t,s) = \partial_t^2   \phi_1^{(4)}(t,s)  $ and $	\phi_4^{(4)}(t,s) = \partial_t^3   \phi_1^{(4)}(t,s)  $. 

Thus, the group action from $X_4$ shifts $t$ in a specific way. 

\section{Two-dimensional first order representations}	

We now generalise the previous discussion to a more generic scenario and explore systematically some specific maps of the equations of motion resulting from the Lagrangian density ${\cal L}(q,\dot{q} , \ddot{q})$ in (\ref{LPU}) to those arising from a two dimensional Lagrangian involving at most first order time derivatives 
\begin{equation}
	{\cal L}(x,y,\dot{x}, \dot{y}) = \frac{a_x}{2} \dot{x}^2 +  \frac{a_y}{2} \dot{y}^2 -  \frac{b_x}{2} x^2 -  \frac{b_y}{2} y^2 - g x y , \qquad a_x,a_y,b_x,b_y,g\in \mathbb{R} .  \label{genLxy}
\end{equation}
The transformations are assumed to be of the form
\begin{equation}
	x(t):= \mu_0 q(t) +\mu_1 \dot{q}(t) +   \mu_2 \ddot{q}(t), \quad \text{and } \quad   
	y(t):= \nu_0 q(t) +\nu_1 \dot{q}(t) +   \nu_2 \ddot{q}(t).  \label{pointt}
\end{equation}
with real constants $\mu_i$, $\nu_i$, $i=0,1,2$ that need to be determined. Being two dimensional, the Lagrangian density ${\cal L}(x,y,\dot{x}, \dot{y}) $ will give rise to two coupled second order equations of motion
\begin{equation}
	a_x \ddot{x}  + b_x x + g y =0, \qquad \text{and} \qquad  a_y \ddot{y}  + b_y y + g x =0.  \label{equnmxy}
\end{equation}
Next we make contact with the PU fourth order equation of motion in (\ref{equm1}). We identify four distinct cases corresponding to two scenarios: either both equations in (\ref{equnmxy}) transform into the PU-equation in (\ref{equm1}), or only the first equation (\ref{equnmxy}) transforms into the PU-equation in (\ref{equm1}) while the second equation trivially vanishes. We denote the first type of transformation as Ta and the second as Tb. 

Using the abbreviations $\rho^\pm_g := \pm \sqrt{\alpha^2-4\beta-\frac{4g^2}{a_x a_y}}$,  $\rho_0^\pm=\rho^\pm_{g=0} $, $\tau := b_x^2 - a_x b_x \alpha + a_x^2 \beta $ the four solutions are:

\noindent
\textbf{Ta1$^\pm$:}
\begin{equation}
	\begin{aligned}
 a_x &\neq 0 , \quad b_x^\pm = \frac{a_x}{2} \left(\alpha - \frac{2g}{a_y} + \rho^\pm_0  \right) , \quad
 \mu_0^\pm = \frac{1}{2 a_x} \left(   \alpha + \rho^\mp_0  \right) ,  \quad \mu_1 = 0 , \quad \mu_2 = \frac{1}{a_x},  \\
 a_y &\neq 0 , \quad b_y^\pm = \frac{a_y}{2} \left(\alpha - \frac{2g}{a_x} + \rho^\pm_0  \right) , \quad
 \nu_0^\pm = \frac{1}{2 a_y} \left(   \alpha + \rho^\mp_0  \right) ,  \quad \nu_1 = 0 , \quad \nu_2 = \frac{1}{a_y},
 	\end{aligned} \label{Ta1tran}
\end{equation}

\noindent
\textbf{Ta2$^\pm$:}
\begin{equation}
	\begin{aligned}
		a_x &\neq 0 , \quad b_x^\pm = \frac{a_x}{2} \left(\alpha + \rho^\pm_g \right)  , \quad
		\mu_0^\pm = \frac{1}{2 a_x} \left(   \alpha + \rho^\mp_g -\frac{2 g}{a_y} \right) ,  \quad \mu_1 = 0 , \quad \mu_2 = \frac{1}{a_x},  \\
		a_y &\neq 0 , \quad b_y^\pm = \frac{a_y}{2} \left(\alpha + \rho^\mp_g \right) , \quad
		\nu_0^\pm = \frac{1}{2 a_y} \left(   \alpha + \rho^\pm_g -\frac{2 g}{a_x} \right) ,  \quad \nu_1 = 0 , \quad \nu_2 = \frac{1}{a_y}, 
	\end{aligned} \label{Ta2tran}
\end{equation}

\noindent
\textbf{Tb1:}   
\begin{equation}
	\begin{aligned}
		a_x &\neq 0 , \qquad \quad \, \, b_x \neq \frac{a_x}{2} \left(\alpha + \rho^\pm_0 \right)  , \quad \,
		\mu_0= \frac{1}{ a_x} \left(   \alpha - \frac{b_x}{a_x} \right) ,  \quad \mu_1 = 0 , \quad \mu_2 = \frac{1}{a_x},  \\
		a_y &= -\frac{ a_x g^2}{ \tau }, \quad 
		b_y = \frac{g^2 (b_x -a_x \alpha)   }{\tau }, \quad
		\nu_0 = \frac{\tau}{g a_x^2}  ,  \quad \nu_1 = 0 , \quad \nu_2 = 0 , \quad
		g \neq 0, 
	\end{aligned}  \label{Tb1tran}
\end{equation}

\noindent
\textbf{Tb2$^\pm$:}
\begin{equation}
	\begin{aligned}
		a_x &\neq 0 ,   \quad b_x^\pm =  \frac{g^2}{b_y} +\frac{a_x}{2} \left(\alpha + \rho_0^\pm \right)  , \quad
		\mu_0^\pm =   \frac{2 \beta}{a_x \left( \alpha + \rho^\pm_0       \right)  }      ,  \quad \mu_1 = 0 , \quad \mu_2 = \frac{1}{a_x},  \\
		a_y &=0 , \quad b_y^\pm \neq 0 , \quad
		\nu_0^\pm =  \frac{2 \beta g }{a_x b_y \left( \alpha + \rho^\pm_0       \right)  }   ,  \quad \nu_1 = 0 , \quad \nu_2 = -\frac{g}{a_x b_y}.
	\end{aligned}  \label{Tb2tran}
\end{equation}
The most relevant solutions are Ta2$^\pm$ and Tb1, but we report here all solutions for completeness. We observe for instance that Ta1$^\pm$ implies that $x \propto y$, which is not a very appealing features from the start. 

\subsection{Transformed Hamiltonians}
Having obtained several versions of the transformed Lagrangian in form of (\ref{genLxy}), we carry out a Legendre transformation to obtain the corresponding Hamiltonians. We identify here $p_x = a_x \dot{x}$, $p_y = a_y \dot{y}$ as momenta and compute
\begin{equation}
{\cal H} (x,y,\dot{x}, \dot{y})  = p_x \dot{x} +p_y \dot{y} - {\cal L}(x,y,\dot{x}, \dot{y}) .   \label{Legtr}
\end{equation}
Translating back to the $\vec{q}$-variables we find the following solutions for each of the relevant transformations
\begin{eqnarray}
	  {\cal H}_{\text{Ta1}^\pm}(q, \dot{q}, \ddot{q}, \dddot{q}) &=&- \frac{(a_x+a_y)}{a_x a_y} 
	  \left[ M^\pm(\omega_1^2 ,\omega_2^2 ) {\cal H} _{1} + {\cal H} _{2} \right],   \label{HT1} \\
	 {\cal H}_{\text{Ta2}^\pm}(q, \dot{q}, \ddot{q}, \dddot{q}) &=& 
\frac{1}{2 a_x a_y} \left\{	[4g-\rho^\pm_g (a_x-a_y) -\alpha (a_x+a_y)] {\cal H}_1 -2(a_x+a_y)	{\cal H}_2 \right\}, \quad  \label{HT2} \quad  \\
 {\cal H}_{\text{Tb1}}(q, \dot{q}, \ddot{q}, \dddot{q}) &=&
\frac{1}{a_x} \left[\left(\frac{b_x}{a_x}-\alpha \right) {\cal H}_{1} - {\cal H}_{2} \right],  \label{HT3}\\
 {\cal H}_{\text{Tb2}^\pm}(q, \dot{q}, \ddot{q}, \dddot{q}) &=& -
\frac{1}{a_x} \left[ M^\pm(\omega_1^2 ,\omega_2^2 ) {\cal H}_{1} + {\cal H}_{2} \right],  \label{HT4}
\end{eqnarray}
where we introduced the functions $M^+(x,y) := \min(x,y)$, $M^-(x,y) := \max(x,y)$ and ${\cal H}_{1}$, ${\cal H}_{2}$ are defined in (\ref{Ham1}) and (\ref{Ham2}), respectively.

\subsection{Flow preserving Poisson brackets}
Next we address the question of which of the transformed Hamiltonians preserves the flow of the original system and identify the corresponding Poisson bracket structures. In addition, we consider which of the Hamiltonians allows for the preservation of the Poisson bracket structure for either of the two Bi-Hamiltonians. 

Comparing with (\ref{barJH}) we note that all transformed Hamiltonians are already in the form of $\bar{ {\cal H}}$ with different coefficients $ c_3$ and $ c_4$. Thus we simply have to solve (\ref{a1234}) for the coefficients $ c_1$ and $ c_2$ to identify the corresponding flow preserving Poisson bracket structure. We find
\begin{equation}
	J_T = \frac{ c_3}{( c_3 -  c_4 \omega_1^2) ( c_3 -  c_4 \omega_2^2)} J_1 + 
	\frac{ c_4  \omega_1^2  \omega_2^2   }{( c_3 -  c_4 \omega_1^2) ( c_3 -  c_4 \omega_2^2)} J_2,
	\quad  c_3 \neq  c_4  \omega_1^2,  \,\,   c_3 \neq  c_4  \omega_2^2 .	 \label{Jflow}
\end{equation}
Thus $ {\cal H}_{Ta1^\pm}$ and $ {\cal H}_{Tb2^\pm}$ do not possess a Poisson bracket structure that preserve the original flow as in these cases the coefficients become singular. In turn, for the other two cases we find perfectly well-defined flow preserving Poisson bracket tensors
\begin{equation}
		\begin{aligned}
	J_{\text{Ta2}^\pm} &= \frac{a_x^2 a_y^2 }{ 2\left[ a_y g - a_x (g + a_y \rho^\pm_g)    \right]^2 } \left\{  \left[
	a_x (\alpha + \rho^\pm_g) + a_y (\alpha - \rho^\pm_g)  -4g	\right]   J_1 +
	2  \beta (a_x+a_y) J_2   \right\} , \quad
		      \\
	J_{\text{Tb1}} &= \frac{a_x^2}{ b_x^2 - \alpha a_x b_x  + \beta a_x^2 }  
	\left[        \left(  b_x-   \alpha  a_x           \right) J_1 -  \beta a_x  J_2    \right] ,   \label{J2}
		\end{aligned}
\end{equation}
Moreover, we can of course achieve to keep original Poisson bracket structures most commonly used for specific choices of the free parameters left in our analysis. We have
 $J_{\text{Ta2}^\pm} = J_1$ for the choice $a_x =-a_y= \pm \sqrt{\alpha^2 -4 \beta - 4g }$ when $\alpha^2 -4 > 4g $ and $J_{\text{Ta2}^\pm} = J_2$ for instance for the choice $a_x=1$, $a_y=-1/2$ and $c= -\alpha \pm 3 \sqrt{\beta} /\sqrt{2}$. For the Tb1 transformation we can only maintain the $J_2$ Poisson bracket structure, we have  $J_{\text{Tb1}} = J_2$ for $a_x=-1$ and $b_x=- \alpha$.
 
 Let us now see how these brackets relations translate to the standard $\vec{x} = \{x,y,p_x,p_y \}$ variables. Implementing the general transformation (\ref{pointt}) with $\mu_1=\nu_1=0$, as this holds for all transformations, on
 \begin{equation}
 	\left\{ F,G  \right\}  = \nabla_q F(\vec{q}) \bar{J}^q  \nabla_q G(\vec{q})  = \nabla_x F(\vec{x})  \bar{J}^x  \nabla_x G(\vec{x}) , \quad \text{with} \quad
 	\bar{J}^x_{ij} =   \sum_{n,m=1}^4  \frac{\partial x_i}{ \partial q_n} \bar{J}^q_{nm}   \frac{\partial x_j}{ \partial q_m}  
 \end{equation}
 we find the only nonvanishing Poisson brackets in the new variables are 
 \begin{eqnarray}
 	\left\{ x,p_x     \right\} &=& a_x \left[ \mu _2^2 \left(\alpha   c_1- c_2\right)+\frac{ c_2 \mu _0^2}{\beta }-2  c_1 \mu _2 \mu _0\right], \label{nanvP1} \\
 	\left\{ x,p_y     \right\} &=&  \mu _2 a_y \left[ \nu _2 \left( c_2-\alpha   c_1\right)+ c_1 \nu _0\right]+\mu _0 a_y\left( c_1 \nu _2 -\frac{ c_2 \nu _0 }{\beta }\right) ,  \\
 	\left\{ y,p_x     \right\} &=& \mu _2 a_x \left[ \nu _2 \left( c_2-\alpha   c_1\right)+ c_1 \nu _0\right] + \mu _0 a_x \left(\frac{ c_2 \nu _0 }{\beta }- c_1 \nu _2 \right), \\
 	\left\{ y,p_y     \right\} &=& a_y \left[\frac{ c_2 \nu _0^2}{\beta }-\nu _2^2 \left( c_2-\alpha   c_1\right)-2  c_1 \nu _2 \nu _0\right].  \label{nanvP2}
 \end{eqnarray}
 For the transformations Ta2$^\pm$ and Tb1 with our specific solutions for the Hamiltonian (\ref{HT2}) and  (\ref{HT3}), these relations reduce automatically to the canonical Poisson bracket relations simply by substituting the corresponding solutions from  (\ref{Ta2tran})  and (\ref{Tb1tran}), respectively,
 \begin{equation}
 	\left\{ x,p_x     \right\} =1, \quad 	\left\{ x,p_y     \right\} =0, \quad  	\left\{ y,p_x     \right\} =0, \quad  	\left\{ y,p_y     \right\} =1.
 \end{equation}
 The parameters that are free in these transformations are left arbitrary. Starting from the Legendre transformed Lagrangian (\ref{Legtr}) and solely working with $\vec{x}$ this is of course expected, so that the identities for the right hand sides of (\ref{nanvP1})-(\ref{nanvP2}) constitute a nontrivial check for our solutions.
 
\subsection{Ghostly models, coupled oscillators with Lorentzian kinetic term}
The standard transformation, already known to Pais and Uhlenbeck \cite{pais1950field}, maps  to a system of two harmonic oscillators of opposite signs. This scenario can be obtained from the solution Ta2$^\pm$ with $g=0$ and $a_x = - a_y$. With $\alpha$ and $\beta$ expressed in terms of the two frequencies and for definiteness $\omega_1^2> \omega_2^2 $ relation (\ref{HT2}) reduces to 
\begin{equation}
	 {\cal H}_{\text{Ta2}^+}(q, \dot{q}, \ddot{q}, \dddot{q}) =   \frac{\omega_2^2 - \omega_1^2}{a_y}   {\cal H}_1 .
\end{equation}
As in (\ref{pointt}) the coefficients $\mu_1$ and $\nu_1$ are always zero we can express the PU-variables uniquely in terms of the transformed variables
\begin{equation}
	q = \frac{\mu_2 y-\nu_2 x}{\mu_2 \nu_0 - \mu_0 \nu_2}, \,\,
	\dot{q} = \frac{ a_x \mu_2 p_y - a_y \nu_2 p_x  }{ a_x a_y (\mu_2 \nu_0 - \mu_0 \nu_2)}, \,\,
	\ddot{q} = \frac{ \nu_0 x - \mu_0 y }{\mu_2 \nu_0 - \mu_0 \nu_2}, \,\,
	\dddot{q} = \frac{ a_y \nu_0 p_x - a_x \mu_0 p_y }{ a_x a_y (  \mu_2 \nu_0 - \mu_0 \nu_2 )}.   \label{qqqxy}
\end{equation}
We notice that these relations do not hold for Ta1$^\pm$ as in that case $\mu_2 \nu_0 = \mu_0 \nu_2 $. Taking $a_y=-1$ we obtain directly
\begin{equation}
	{\cal H}_{\text{Ta2}^+}(x,y,p_x,p_y) =  \frac{1}{2} \left( p_x^2 + \omega_1 x^2  \right)- \frac{1}{2} \left( p_y^2 + \omega_2 y^2  \right) ,
\end{equation}
which evidently is not positive definite with a quantum spectrum not bounded from below. Keeping $g \neq 0$ leads to a ghostly system in form of coupled oscillators with Lorentzian kinetic term
\begin{equation}
	{\cal H}'_{\text{Ta2}^\pm}(x,y,p_x,p_y) =  \frac{1}{2} \left( p_x^2  -p_y^2  \right)  + \frac{1}{4} \left( \rho^\pm_g + \omega_1^2  +  \omega_2^2  \right) x^2 + \frac{1}{4} \left( \rho^\pm_g - \omega_1^2  -  \omega_2^2  \right) y^2 + g x y. 
\end{equation}
The model is a specific case of a class of systems whose classical version was recently introduced in \cite{diez2024foundations}. It was quantised in \cite{fring2025quantisations} and a classical field theoretical version was studied in \cite{deffayet2025ghostly}. The standard Hamiltonian is recovered when setting $g$ to zero, i.e. we have $\lim_{g \rightarrow 0} 	{\cal H}'_{\text{Ta2}^+} = {\cal H}_{\text{Ta2}^+}$. Focussing on the dynamics of these systems, their ghostly nature can be removed by invoking their Bi-Hamiltonian structure.

Taking instead $a_y=1$ the system (\ref{HT2}) reduces to two positive definite space coupled harmonic oscillators
\begin{equation}
	\check{{\cal H}}_{\text{Ta2}^\pm}(x,y,p_x,p_y) =  \frac{1}{2} \left( p_x^2 + p_y^2  \right)  + \frac{1}{4} \left( \rho^\pm_g + \omega_1^2  +  \omega_2^2  \right) x^2 + \frac{1}{4} \left( \rho^\mp_g + \omega_1^2  +  \omega_2^2  \right) y^2 + g x y,
\end{equation}
resembling a system recently studied in \cite{bend2016analytic,felski2018analytic}.

\section{Positive definite versions of the PU model}
Next, we address the crucial question of whether it is possible to use any of the transformations in which the conversion to the form (\ref{H12H21}) is positive definite, i.e. for which the inequalities (\ref{PDC}) hold. This is indeed possible. Taking for simplicity $a_x=a_y=1$, we have 
\begin{equation}
	{\cal H}_{Ta2} =H_{12}^{a2} +H_{21}^{a2}, \qquad          H_{ij}^{a2} =  \frac{2 g+\omega _i^2-\omega _j^2}{2 \omega _i^2-2 \omega _j^2} \left[  \left(  \dddot{q} + \omega_j^2 \dot{q}    \right)^2  +  \omega_i^2  \left(  \ddot{q} + \omega_j^2 q    \right)^2      \right],
\end{equation}
which is positive definite iff $ \omega _2^2-\omega _1^2  <2 g < \omega _1^2-\omega _2^2 $ or $ \omega _1^2-\omega _2^2  <2 g < \omega _2^2-\omega _1^2 $ and $ \omega_1 \neq 0$, $ \omega_2 \neq 0$. Likewise, fixing only $a_x=1$ for simplicity, we obtain a positive definite Hamiltonian from the transformation Tb1
\begin{equation}
	{\cal H}_{Tb1} =H_{12}^{b1} +H_{21}^{b1}, \qquad          H_{ij}^{b1} = \frac{b_x - \omega_j }{2 \omega _i^2-2 \omega _j^2}    \left[  \left(  \dddot{q} + \omega_j^2 \dot{q}    \right)^2  +  \omega_i^2  \left(  \ddot{q} + \omega_j^2 q    \right)^2      \right],
\end{equation}
when $ \omega_1^2 < b_x < \omega_2^2  $ or $ \omega_2^2 < b_x < \omega_1^2  $. Using (\ref{qqqxy}), we transform the Hamiltonians to the $\vec{x}$-variables
\begin{eqnarray}
		{\cal H}_{Ta2} &=&H_{12}^{a2} +H_{21}^{a2}, \quad          H_{ij}^{a2} =  \frac{2 g+\omega _i^2-\omega _j^2}{2 \omega _i^2-2 \omega _j^2} \left[ (p_x \kappa_+^{ij} +  p_y \kappa_-^{ij} )^2     + \omega_i^2 (x \kappa_+^{ij} +  y \kappa_-^{ij} )^2        \right], \qquad \\
			{\cal H}_{Tb1} &=&H_{12}^{b1} +H_{21}^{b1}, \quad          H_{ij}^{b1} = \frac{b_x - \omega_j }{2 \omega _i^2-2 \omega _j^2}   \left[   (  p_x \lambda_\nu^j + p_y \tau \lambda_\mu^j           )^2            + \omega_i^2 \left( x \lambda_\nu^j + y \lambda_\mu^j     \right)^2             \right],
\end{eqnarray}
with  $\tau= (b_x-\omega_1^2)(b_x-\omega_2^2)/g^2$ and 
\begin{equation}
 \kappa_{\pm} ^{ij}= \frac{1}{2} \pm \frac{  \rho^{\pm}_g }{4 g + 2 \omega_i^2  - 2 \omega_j^2}, \quad
 \lambda_\mu^i = \frac{\mu_0 - \mu_2 \omega_i^2}{\mu_2 \nu_0 - \mu_0 \nu_2 }, \quad
  \lambda_\nu^i = \frac{\nu_0 - \nu_2 \omega_i^2}{\mu_2 \nu_0 - \mu_0 \nu_2 }. 
\end{equation}

 A positive definite version of the PU model was previously proposed in \cite{stephen2008ostro,most2010h} without the associated Poisson bracket structure. Let us see how the suggested version fits as special case into the above scheme. The proposed Hamiltonian in the notation of \cite{most2010h} is 
 \begin{equation}
     	{\cal H}_{SM}(w,z,p_w,p_z)= \frac{p_w^2}{2 \mu_w} +\frac{p_z^2}{2 \mu_z}  +\left( \nu_w  w - \frac{\nu_z \Omega^2 }{\sqrt{\alpha^2 -\delta}}  z    \right) + \frac{4 \beta \nu_z^2}{\alpha^2 - \delta} z^2,
 \end{equation}
where $\alpha$ and $\beta$ play the same role as here and $\delta = \alpha^2 -4 \beta -\Omega^4$, $\nu_w = \sqrt{ \mu_w( \alpha \pm\sqrt{\delta})}/2$, $\nu_z = \sqrt{ \mu_z( \alpha \mp\sqrt{\delta})}/2$ and $\Omega= (4 \mu_z/\mu_w)^{1/4} \tau^{-1}$. The coordinates are related to the PU-variable as $z(t)= q(t)$, $w(t)= \lambda \tau^2 q(t) + \tau^2 \ddot{q}(t)$, $p_w(t)=\mu_w \dot{w}(t)$, $p_z(t)=\mu_z \dot{z}(t)$ with $\lambda= (\alpha \pm \sqrt{\delta})/2$. Identifying $w\equiv x$, $z\equiv y$ this transformation simply corresponds to Tb1 with $\nu_0=1$, $\mu_0 = \lambda \tau^2$, $\mu_2= \tau^{-2}$, such that
 \begin{equation}
    a_x=\tau^{-2},  \quad  a_y= \frac{\mu_z}{\mu_w \tau^2}, \quad  b_x= \frac{\alpha -\lambda}{\tau^2} \quad  b_y= \frac{\mu_z \lambda}{\mu_w \tau^2}, \quad g=-\frac{\Omega^4}{4}.  \label{Alicoeff}
\end{equation}
The appropriate Poisson bracket tensor is now simply (\ref{J2}) with (\ref{Alicoeff}). Crucially one needs to use the Bi-Hamiltonian structure to achieve that.

 Another positive definite version was proposed in \cite{nucci1,nucci2,nucci3} constructed from the sum of the squares of four Noether currents, see expression for $I_{\text{aut}}$ in equation (20) of \cite{nucci1}. Compared with our expressions we find
\begin{equation}
         I_{\text{aut}} = -2 \alpha {\cal H}_1 + 4 {\cal H}_2 .
\end{equation}	
It follows from (\ref{Lieconst}) that in order to preserve the original flow for this combination we need to set 
 $c_1 = - \frac{ \alpha}{ 6 \alpha^2 + 8 \beta} $ and $c_2 =  \frac{\beta}{ 3 \alpha^2 + 4 \beta} $ in the linear combination of the Poisson tensor. It is also clear from this observation that the analysis provided \cite{nucci1,nucci2,nucci3} in incomplete. 

\subsection{PU model with interaction terms}
There have been a number of studies considering the PU model with an additional interaction terms \cite{pav1pu,pav2pu,pav3pu}, mainly to investigate whether the solutions remain stable in those circumstances. Here we see that in that situation the Bi-Hamiltonian structure is destroyed, so that one can no longer exploit it to find equivalent representations. We consider
\begin{equation}
	{\cal H}_1^{\text{int}} \left( q, \dot{q}, \ddot{q} , \dddot{q}  \right) = 	{\cal H}_1 \left( q, \dot{q}, \ddot{q} , \dddot{q}  \right) +  V(q), 
\end{equation}	 	
where $V(q)$ is an arbitrary potential. In the corresponding flow equation we have to replace the vector field in  (\ref{1flow}) by 
\begin{equation}
	 V_1^{\text{int}}= \sum_{i=1}^4 v_i \partial_{q_i}   = \dot{q} \partial_q +  \ddot{q} \partial_{\dot{q}} 
	+  \dddot{q} \partial_{\ddot{q}} - \left[ \alpha \ddot{q} + \beta q - V'(q)   \right] \partial_{ \dddot{q} }.
\end{equation}
It is then easily verified that the only compatible solution to $J \nabla 	{\cal H}_1^{\text{int}}  =  V_1^{\text{int}}(\vec{q}) $ is $J_1$.

Similarly we define
\begin{equation}
	{\cal H}_2^{\text{int}} \left( q, \dot{q}, \ddot{q} , \dddot{q}  \right) = 	{\cal H}_2 \left( q, \dot{q}, \ddot{q} , \dddot{q}  \right) + W(\ddot{q}),
\end{equation}	 	
where $W(\ddot{q})$ is an arbitrary function of $\ddot{q}$. In the flow equation we have to replace the vector field in  (\ref{1flow}) by 
\begin{equation}
	V_2^{\text{int}}= \sum_{i=1}^4 v_i \partial_{q_i}   = \dot{q} \partial_q +  \ddot{q} \partial_{\dot{q}} 
	+  \dddot{q} \partial_{\ddot{q}} - \left[ \alpha \ddot{q} + \beta q - W'(\ddot{q})   \right] \partial_{ \dddot{q} }.
\end{equation}
It is then easily verified that the only compatible solution to $J \nabla 	{\cal H}_2^{\text{int}}  =  V_2^{\text{int}}(\vec{q}) $ is $J_2$. Notice that $\ddot{q}$ is also a function of $x$ and $y$ in our transformation (\ref{qqqxy}), so that $W(\ddot{q})$ is still a potential in the standard sense that does not depend on the momenta. For the flows would become identical for $V=W$ and $q=\ddot{q}$, but this is not possible as for the $\mu_2 = - \mu_0$, $\nu_2 = - \nu_0$ the transformation becomes singular. Thus the Bi-Hamiltonian structure can not be maintained.

Next we consider how interaction terms will affect the transformation from the fourth order time-derivative theory in one dimension to second order time-derivative theory in two dimensions. Adding a potential term of the form $-V(q)$ to the Lagrangian 	${\cal L}_{\text{PU}}(q,\dot{q} , \ddot{q})$ in (\ref{LPU}) will simply add a term $-V'(q)$ to the equation of motion in (\ref{equm1}). Similarly adding a term $-V(x,y)$ to the Lagrangian in (\ref{genLxy}) will change the corresponding equations of motion to
\begin{equation}
	a_x \ddot{x}  + b_x x + g y  + \partial_ x V(x,y) =0, \qquad \text{and} \qquad  a_y \ddot{y}  + b_y y + g x + \partial_ y V(x,y) =0.  \label{equnmxyint}
\end{equation}
Recalling our two transformation scenarios, Ta and Tb, for mapping these equations to the higher-order equation of motion, only Ta remains viable. In scenario Tb, one of the equations must vanish trivially, which cannot be achieved while preserving linearity in the transformation. For scenario Ta to work, both equations in (\ref{equnmxyint}) must be converted into the PU equation, which imposes the following constraints:
\begin{equation}
	   \partial_ x V(x,y)= \frac{dV}{dq}  \frac{dq}{dx} = - \frac{dV}{dq}, \qquad \text{and}   \qquad 
	   \partial_ y V(x,y)= \frac{dV}{dq}  \frac{dq}{dy} = - \frac{dV}{dq} .
\end{equation}
Thus, with $dq/dx=-1$ and $dq/dy=-1$, we obtain from (\ref{qqqxy}) the constraints 
\begin{equation}
   \frac{\nu_2}{ \mu_2 \nu_0 - \mu_0 \nu_2 } =1 , \qquad \text{and}   \qquad 
- \frac{\mu_2}{ \mu_2 \nu_0 - \mu_0 \nu_2 } =1  .
\end{equation}
For the transformation Ta1$^\pm$, these constraints become singular and are therefore not viable. However, using Ta2$^\pm$yields the solutions
  \begin{equation}
  a_x=-a_y = \pm \sqrt{\alpha^2 - 4 \beta - 4 g}  .
  \end{equation} 
  Thus, it is indeed possible to carry out the transformation for the PU with an interaction term in form of an arbitrary potential of the specified form. A more general investigation of interaction terms of the form $V(q, \dot{q}, \ddot{q}, \dddot{q})$ is left for future work. Evidently, the flow and the Poisson bracket structure are more substantially modified in such a generalized setting.

\section{Conclusion}

We have presented a comprehensive analysis of the PU model, a prototypical higher time-derivative theory, from the perspective of Lie symmetries. By explicitly identifying the Lie symmetries of the fourth-order PU oscillator and leveraging its Bi-Hamiltonian structure, we constructed alternative Poisson bracket formulations that preserve the original dynamics while admitting positive definite Hamiltonians.

Moreover, we systematically explored a variety of transformations that map the fourth-order dynamics of the PU model into equivalent two-dimensional, first-order systems. Among these, we identified transformations that allow for positive definite reformulations and canonical Poisson brackets, thereby enabling stable classical and quantum formulations. We demonstrated that while some well-known ghostly realizations of the PU model arise naturally within our framework, the use of Bi-Hamiltonian structures allows for stable alternatives with equivalent dynamics.

  In summary: Our procedure consist of the following steps: i) Identify the Bi-Hamiltonian (possibly multi-Hamiltonian) structure for the dynamical flow of the higher time-derivative equation of motion. ii) Identify the autonomous Lie symmetries for the vector equation of the dynamical flow. iii) Construct new dynamical flows from linear combinations of the Poisson bracket tensors and the Bi-Hamiltonians. iv) Use the Lie symmetries to identify the constraints that preserve the original flow. v) Select all relevant solutions by imposing positive-definiteness on the Hamiltonian. vi) Map the higher time-derivative theory to coupled lower lower dimensional theories and identify the viable models in those higher dimensional spaces.

Importantly, we showed for our model that the inclusion of interaction terms generally breaks the Bi-Hamiltonian structure, highlighting both the power and limitations of the approach. These findings suggest promising directions for future work, including the construction of stable interacting HTDTs, generalizations to field-theoretic settings, and further investigation into the quantization of positive-definite PU models. Our analysis reinforces the central role of symmetry in navigating the complexities of the dynamics of HTDTs.

 \medskip

\noindent {\bf Acknowledgments}: AlF acknowledges funding from the Max Planck Society’s Lise Meitner Excellence Program 2.0. BT is supported by a City St George's, University of London Research Fellowship.

\newif\ifabfull\abfulltrue

\end{document}